\documentclass[a4paper,11pt]{article}
\usepackage{amssymb}
\usepackage{amsmath}
\usepackage[latin1]{inputenc}
\usepackage{tabularx}
\usepackage{graphicx}
\usepackage{caption}
\usepackage{mathtools}
\usepackage{enumerate}
\usepackage{color}
\usepackage{url}
\usepackage[colorlinks=true,linkcolor=black, citecolor=blue, urlcolor=blue]{hyperref}
\usepackage{dsfont}

\begin{document}


\title{Oscillations and irregular persistent firing patterns in a homogeneous network of excitatory stochastic neurons with gap junctions in the mean-field limit}


\author{G.~Via\thanks{gviarodriguez@gmail.com}\\
{\it Universidade de S\~ao Paulo}}

\maketitle

\providecommand{\keywords}[1]{\textbf{\textit{Keywords---}} #1}




\begin{abstract}
We study systematically a recently developed mathematical model for networks of excitatory stochastic point spiking neurons in the mean-field limit.
The neurons are leaky and connected by uniform connectivity strengths of both passive electrical (gap junctions) and chemical synapses, and they emit a spike
with a firing rate probability that depends on their actual membrane potential.
This allows for a treatable description by collapsing all of the sources of noise in the single firing rate function.
The probability density function of membrane potentials across the population was shown to solve a non-trivial integro-differential equation that reminds of
the advection equation common to fluid dynamics.
We give pseudo-analytical expressions for the non-trivial invariant distributions that solve it (the trivial distribution with no persistent activity and a Dirac delta mass at zero is always solution) and numerical 
solutions of the full time-dependent equation.
These results extend previous results that either considered neurons with leakage and no gap junctions or viceversa.
The non-trivial distributions are of compact support whenever leakage or gap junction intensity are non-zero, otherwise heavy-tailed distributions arise.
As previous studies show the non-trivial distributions can be continuous or discontinuous, depending on the competition of the spiking rate at the highest membrane potential
with positive probability and the combined pull of leakage and gap junctions.
When the latter dominates the discontinuity is infinite, but these solutions seem to be always unstable.
Oscillations of the global activity of the network in the so called synchronous states were found previously for non-leaky neurons.
We show here how they remain when weak leakage is included.
When it is too strong the activity dies out and the system ends up at the trivial invariant distribution.
These results show how a purely excitatory network of stochastic leaky neurons can sustain global oscillations, and how this simple model offers great potential for
large-scale modeling of neural networks.
\end{abstract}









\newpage

\tableofcontents



\section{Introduction}



It is been estimated that the human brain is composed of about $10^{11}$ neurons\cite{herculano}. Yet it is common to study the smaller nervous systems of non-human species, and/or to focus in one or few of the functional and anatomically differentiated structures that compose them. Even in this case the number of neurons remains very high. This makes it unfeasible to perform computational simulations of neural networks of sizes comparable to the real systems at the microscopic scale. It is even harder to give mathematically rigorous treatments of them. For this reason mean-field and other hydrodynamic limits\cite{faugerasTouboul} of neural network models present a very promising tool for the modeling of these systems at the meso- and macroscopic scales. These different scales are typically defined by the nature of the segregated composing units of the network\cite{spornsConnectome}: single neurons in the miscroscopic case, small functional and/or anatomically homogeneous populations (like cortical columns or minicolumns) in the mesoscale, and the larger functional and anatomically differentiated structures in the macroscopic one. In this framework, the hydrodynamic limit often refers to the limit when the number of neurons tends to infinity, while the mean-field limit is the particular case where the connections between all pairs of neurons are of same strength. The mean-field limit, thus, only considers the mean influence on each neuron from its neighbors. Moreover, these approximations assume a high degree of homogeneity in the neuronal and synaptic properties across the population, a sensible constraint in highly homogeneous populations like cortical minicolumns. The approximation of number of neurons $N \rightarrow \infty$ remains useful despite of the finite size of these populations ($N \approx 80-100$).




These models present a great advantage since they allow for the collapse of a large number of variables and parameters into just a few statistical quantities. This allows one to reduce drastically the dimensionality of the system. It is easy to perceive how powerful this can be, allowing for a computational and even sometimes mathematical treatment of problems that were initially out of reach. Thus, it is not surprising that in the recent years the interest on these type of models for the simulation of neural networks has increased significantly (see \cite{gerstnerkistler, faugerasTouboul, fournierLocherbach} and references therein).

One important macroscopic property of neural networks is their capacity to sustain oscillations of the global activity of neurons, where they present a high degree of synchrony\cite{reviewRythms}. This means that the firing times of different neurons in the population present high correlations. These global states are often referred to as synchronous states, in contrast to asynchronous states where these correlations are low. Synchronous and asynchronous states were previously found in neural network models in the mean-field limit, e.g. using the leaky-integrate-and-fire model to describe the activity of the composing neurons\cite{gerstnerkistler, brunel}. In these works the network is often studied in the so called balanced state\cite{sompolinsky}, where the mean excitatory and inhibitory inputs into each neuron cancel each other. This is been proposed to be responsible for the high degrees of variability observed in neural spike trains\cite{gerstnerkistler}.

Theoretical studies have shown inhibition and gap junctions capable to give rise to oscillations in the global activity of neural networks\cite{reviewRythms}, and experiments have shown a direct involvement of the latter in this phenomenon\cite{ExpFuncGJs, reviewPattActDev}.
Actually, even excitation was proved capable of this under certain conditions\cite{reviewRythms}, and more recently astrocytes were shown to take part in this process\cite{astrocytesOscs}.
However, most studies that showed this in gap-junctional networks considered either dendro-dendritic gap junctions between inhibitory interneurons or axo-axonal junctions between excitatory ones (see, e.g. \cite{GJinh, gapSynch2, axonalGapSynchTraub}).
To the best of our knowledge, no study considered the effects of the dendro-dendritic coupling in excitatory neurons.
Maybe because it is very rare in the adult mammalian brain\cite{reviewRythms}.
However, they are more prevalent in the developing brain\cite{reviewRythms}, where the synchronized activity of neural populations was shown to play an important role in the development of the networks\cite{correlActDevCircs}.
Moreover, in this phase GABAergic synapses, which become inhibitory in adulthood, are depolarizing\cite{reviewPattActDev}.
Therefore, this type of coupling between excitatory neurons could play a role in the development of the circuitry of the mammalian nervous systems.
In this work we analyze a homogeneous population of excitatory stochastic spiking neurons connected both by chemical and electrical synapses, as well as leakage of ions across their membranes. Our gap junctions simulate dendro-dendritic electrical coupling and neglect the effects of action potentials on neighboring neurons.




The model was shown in \cite{drogoulVeltz} to present a synchronous state, characterized by global oscillations in the network activity, when no leakage is present and the gap junctions are permeable enough. Otherwise, the network can also remain in an asynchronous or a \emph{dead} state. In the latter case the network presents no activity and all neurons remain with zero membrane potential. It is important to recall, though, that when no external input is fed to the network, as is the case in the present work, the sustained activity should fade away in a finite amount of time when finite networks are considered\cite{robertTouboul, drogoulVeltz}. In this sense, our mean-field limit results recover a long transient of the network.

Here we focus on the study of the equations describing a homogeneous neural network in the mean-field limit. For this we use a model based on that introduced in \cite{galveslocherbach}. This work introduced a new class of models for the simulation of neurons and neural networks as stochastic processes within the group of so called \emph{piecewise-deterministic Markovian processes}. It considered the discrete-time evolution of the state of the neural network, while a continuous-time version was derived in \cite{duarteOst}. The solutions to the mean-field limit equations for the discrete-time version of the model with leakage and no gap junctions were studied in \cite{brochiniDiscGL}, and here we consider the mean-field limit equations derived in \cite{deMasi} for the continuous-time version of the model. We consider a slightly more general case with uniform but arbitrary chemical and electrical synaptic strengths and leakage. The mathematical model is as follows. Each neuron in the network can spike with a spontaneous rate that depends on its actual membrane potential. It is thus an escape rate model that concentrates all the sources of noise on the neural dynamics in a single function\cite{gerstnerkistler}. When a spike is triggered, the membrane potential of all neurons connected to it by chemical synapses is increased by an amount given by the strength of the shared connection. At the same time, the membrane potential of the spiking neuron is reset to a reference value of zero. Between spikes, the membrane potential of all neurons drifts deterministically due to the combination of two processes: leakage, which pulls the membrane potential towards zero; and gap junctions, which pulls it towards its mean value across the population. These processes are due to the exchange of ions between the neuron intra- and extracellular spaces across passive leakage channels in the membrane, and to the exchange of ions between the intracellular spaces of pairs of neurons connected by gap junctions.

The present study complements these in \cite{robertTouboul}, where the case with leakage but no gap junctions is considered, and those in \cite{fournierLocherbach} and \cite{drogoulVeltz}, where the opposite case, with gap junctions but no leakage, was analyzed. In particular, we introduce the full model in section \ref{model} and the methods used to solve the equation in section \ref{methods}. The results are divided into two sections. First, in section \ref{stationary}, the invariant distributions of the system are analyzed. Second, section \ref{stability} comprises a computational study of the stability of these distributions. We conclude with a brief summary and some remarks in section \ref{conclusions}.

\section{Model}
\label{model}









As introduced in \cite{duarteOst}, we consider here a network of $N \in \mathbb{N}$ identical neurons which undergo a passive leakage of ions with intensity $\alpha \in \mathbb{R}^+$ across their membranes and are connected all-to-all by chemical synapses of the same strength $W/N$ (mean-field limit), with $W \in \mathbb{R}^+$, and by passive gap junctions of strength $\lambda \in \mathbb{R}^+$. The network is, thus, composed only of excitatory neurons. The gap junctions are assumed to be dendritic and their influence during action potentials is neglected. Axonal and dendritic delays are also neglected. Each neuron fires a spike randomly with instantaneous firing rate $\phi(V)$ dependent on the present membrane potential $V \in \mathbb{R}^+$, where we will consider power firing functions $\phi(V) = (\gamma x)^n$ with gain $\gamma \in \mathbb{R}^+$ and power $n \in \mathbb{N}$. In the finite size model the membrane potential of each neuron follows a jump process where transitions occur when either the neuron receives a spike from another neuron, with a jump of size $W/N$ in the membrane potential, or when the neuron itself spikes, which resets its membrane potential to a reference value $0$.  The spikes of each neuron occur with an instantaneous rate $\phi(V)$ and hence dependent on the present value of membrane potential. From the biological point of view it is sensible to use a non-decreasing $\phi(V)$, which can be chosen in order to fit experimental data. In the absence of spikes the membrane potentials undergo a deterministic drift resulting from the superposition of leakage, that pulls the membrane potential towards $0$ with rate $\alpha$, and of gap junctions, that pull it towards the population mean potential, $\bar{V}$, with rate $\lambda$.

Here the hydrodynamic limit $N \rightarrow \infty$ is considered. By using homogeneous properties for the network (i.e. leakage and connectivity parameters) we are taking into account the effects of averaged quantities over the population, i.e. we consider the effects of the \emph{mean field} on each of the neurons\cite{fournierLocherbach}. We are considering the diffusive limit of weak synapses since the limit $N \rightarrow \infty$ implies $W/N \rightarrow 0$\cite{gerstnerkistler}.

The \emph{empirical measure} corresponding to the vector of membrane potentials was proven in \cite{deMasi} to converge to a deterministic time-dependent limit density $p(V,t){\rm d}V$ of membrane potentials $V$ that solves the non-local transport equation
\begin{equation}
\frac{\partial p}{\partial t} + \frac{\partial ( c(V,t) p )}{\partial V} = -\phi(V) p,
\label{diffEq}
\end{equation}
where
\begin{equation}
c(V,t) = -\alpha V - \lambda (V-\bar{V}(t)) + W \rho(t)
\label{cVt}
\end{equation}
includes the effects of leakage, gap junctions, and chemical synapses, respectively, on the membrane potential of each neuron, and the term $-\phi(V) p$ represents the loss of mass undergone by the system at each value $V$ due to spiking. Here we follow the notation in \cite{brochiniDiscGL}, where the discrete-time version of the system is considered. This equation reminds of the advection equation common to fluid dynamics, with $c(V,t)$ playing the role of the local spontaneous velocity of the fluid and the inhomogeneous term that of a sink where mass is lost. However, the definitions of the mean membrane potential and mean spiking rate,
\begin{eqnarray}
\bar{V}(t) = \int_0^\infty{V p(V,t) {\rm d}V} & \mbox{ and } &
\rho(t) = \int_0^\infty{\phi(V) p(V,t) {\rm d}V},
\label{pnxDef}
\end{eqnarray}
respectively, render it a non-trivial integro-differential equation. Note that $\phi(V)$ is the usual instantaneous firing rate, i.e. for a neuron whose trajectory $V(t)$ is known, $\int_t^{t+\Delta t}{\phi(V(t')){\rm d}t'}$ corresponds to the net probability for that neuron to spike in the interval $\left[t, t+\Delta t\right]$. Then, according to definition \ref{pnxDef}, $\int_t^{t+\Delta t}{\rho(t'){\rm d}t'}$ can be interpreted as the fraction of neurons in the population that spike in that interval, and hence $\rho(t)$ as an instantaneous fraction of spiking neurons per unit time. For this reason, we will often call it \emph{activity} of the network. Moreover, the above equations need to be complemented with the normalization condition
\begin{equation}
\int_0^\infty{p(V,t) {\rm d}V} = 1
\label{norm}
\end{equation}
and the initial and boundary conditions
\begin{eqnarray}
p(V,t=0) = p_0(V) \hspace{1 cm} \mbox{ and } \hspace{1 cm}
p(V=0,t) = \frac{\rho(t)}{\lambda \bar{V}(t) + W \rho(t)},
\label{extConds}
\end{eqnarray}
with some given distribution $p_0(x)$. The latter condition grants the fulfillment of \ref{norm} by adding a point source at $V=0$ that compensates the loss due to the \emph{sink} term. Therefore, this implements the reset potential condition.

The mean-field limit behavior of the discrete-time version of the model was studied in \cite{brochiniDiscGL} for a homogeneous network of neurons that present a passive leakage of ions but no gap junctions at their membranes. In that case the membrane potential distribution was given by the superposition of Dirac delta masses, each containing all neurons that spiked at same time step. These masses advanced as a result of leakage and presynaptic input from other neurons, while their total mass decayed due to spiking of their composing neurons. However, the flux of neurons from one \emph{packet} to another one was not possible. Besides giving a continuous distribution of membrane potentials, our continuous-time version is qualitatively different as exchanges of neurons between packets occurs all the time due to the non-uniform pulls resulting from leakage and gap junctions whenever $\alpha>0$ or $\lambda>0$.

As the studies in \cite{robertTouboul} suggest, where $\alpha=1$, $\lambda=0$ and i.i.d. synaptic strengths are considered, the behavior of the system depends on quantities normalized to $(\gamma W)^n$, which allows one to fully characterize the state of the system in terms of $V/W$, $t (\gamma W)^n$, $\alpha/(W \gamma)^n$, $\lambda/(W \gamma)^n$, $n$, $\rho(t)/(\gamma W)^n$ and $\bar{V}(t)/W$.


The choice for the spiking rate function $\phi(V) = (\gamma V)^n$ allows one to compare the behavior with the deterministic case, where neurons fire deterministically when their membrane potential overcomes a fixed threshold $1/\gamma$, by taking the limit $n \rightarrow \infty$. Moreover, high powers $n$ are closer to the typically used exponential dependence \cite{gerstnerkistler} and may fit better the input-output characteristics of experimental conditions. Networks of deterministic neurons of this type have already been thoroughly studied in the mean-field and other hydrodynamic limits\cite{gerstnerkistler, faugerasTouboul}. However, most of these studies also consider the diffusive limit but with non-uniform synaptic strength distributions and most of the times with inhomogeneous populations of at least two types of neurons, inhibitory and excitatory, in the so called balanced state\cite{sompolinsky}. In this state the mean excitatory and inhibitory inputs into each neuron cancel each other. Thus, we cannot compare our studies in this limit with previous results. This was proposed as a mechanism to explain the high degree of variability observed experimentally in neural spike trains. In this regime different steady states were found including synchronous and asynchronous states, where the spike times of different neurons present high and low correlations, respectively \cite{brunel}. Until recently, only asynchronous states were found for the system we study here, but in \cite{drogoulVeltz} it was shown how a stable synchronous state also arises when gap junctions of sufficient strength are present and $n\gg 1$ (and thus with more realistic properties). It is important to note that these stable states, both synchronous and asynchronous, recover a long transient that fades away after a finite amount of time when $N$ is finite\cite{robertTouboul, drogoulVeltz}. In this case, the activity can only persist for an infinite amount of time if an external input is fed, e.g. $\phi(0)>0$. These studies, altogether with ours, show the rich repertoire of phenomena arising as a result of introducing intrinsic stochasticity into the network composing neurons. It also allows one to prevent unrealistic behavior resulting from the sharp threshold in deterministic models by introducing some of the diverse natural sources of noise in the neuron dynamics in a compact form that allows for mathematical and cheap computational treatments.




Similar models where either leakage or gap junctions are not considered were thoroughly studied previously \cite{deMasi, fournierLocherbach, robertTouboul, drogoulVeltz}. Interestingly, in \cite{robertTouboul} a slightly more general case is considered by taking non-uniform but i.i.d. randomly distributed synaptic strengths that also scale as $\approx 1/N$. However, their results in the hydrodynamic limit prove to depend just on the mean synaptic strength, which allows for a direct comparison with our results. In \cite{duarteOstInhomo} equations were derived for the more general case of inhomogeneous networks with position-dependent leakage intensity, chemical and electrical connection strengths, and spiking rate function, but the behavior of the system was not studied.

In this work we give a systematic study of the homogeneous network but accounting for the leakage across the neurons membranes and both for the chemical and electrical synapses. Due to the impossibility to solve mathematically the full time-dependent integro-differential equation, we give pseudo-analytical expressions for the stationary solutions and solve numerically the full time-dependent equation. Moreover, sharp discontinuities arise both in the invariant and time-dependent $p$ distributions. The additional complications this gives made us restrict our studies to regions of the parameter space where they are not present.


\section{Methods}
\label{methods}

The analytical expressions for the stationary solutions were derived with the aid of \emph{WolframAlpha}\textsuperscript{\textregistered}\cite{wolframAlpha}.


The numerical simulations were performed using finite-difference approximations for the space- and time-derivatives and the time evolution was computed using the Lax-Wendroff method\cite{detailedNotes}. This choice was motivated from the small artificial diffusive effects it introduces in the advection equation, compared to other finite-difference and finite-volume methods.

The code script was adapted from \cite{FSUwebpage} and extended according to \cite{LWnotes}. In particular, we discretise the simulated period of time $t \in \big[0, T\big]$ in $n_t$ time bins and the range of membrane potential values $V \in \left[0, V_\textrm{max} \right]$ in $n_V$ bins, with $V_\textrm{max}$ chosen to be large enough to grant $p(V_\textrm{max},t)<10^{-3}$ for all $t \in \big[0, T\big]$. If this condition was not fulfilled simulations were repeated with larger $V_\textrm{max}$. The density function $p(V,t)$ is computed at the grid points $V \in \{V_0, V_1, ..., V_{n_V}\}$, $t \in \{t_0, t_1, ..., t_{n_t}\}$, with $V_i=i \Delta V$, $t_j = j \Delta t$, with rectangular cells of sides $\Delta t \equiv T/n_t$ and $\Delta V \equiv V_\textrm{max}/n_V$. For constant and homogeneous speed $c$, the method is proven to be stable under the Courant-Fiedrichs-Levy condition, i.e. ${\rm d}t \leq {\rm d}V/C$ \cite{LWnotes}. However, in our case $c=c(V,t)$ is not constant nor uniform. For this reason we choose a time-dependent time step $\Delta t(t) = K \Delta V / \max_{V \in \left[0,V_{\rm max}\right]}{c(V,t)}$, with constant $K<1$ (see Section \ref{stability}). 


The Lax-Wendroff method\cite{LWnotes, detailedNotes} uses a second order Taylor expansion of the derivative $\partial p/\partial t$ to implement a forward-in-time explicit method where $p_{ij}=p(V=V_i,t=t_j)$ is computed from $(p_{ij-1})_{i=0, 1, ..., n_x}$ for each $j>0$ and $i = 1, 2, ..., n_V-1$. At the upper boundary $p_{n_Vj}=0$ $\forall j=0, ..., n_t$ is used, while at the lower edge $p_{0j}$ is computed from the second degree polynomial obtained from the discretised expression for equation \ref{extConds}. For this we use the values $(p_{ij})_{i>0}$ obtained at the present time step. The velocity $c(V_i,t_j)=c_{ij}$ at each grid point is obtained from the averaged quantities $\rho(t_{j-1})=\rho_{j-1}$ and $\bar{V}(t_{j-1})=\bar{V}_{j-1}$ at the previous time step.

For a study of the stability at the non-trivial invariant distributions, all of the computations performed in this work assumed initial distributions $p_0(x) = \left( p_0(x_i) \right)_{i \in 0, 1, ..., n_x}$ given by these distributions, whose pseudo-analytic expression was found previously, as shown in section \ref{stationary}.

It is common to hyperbolic and other convection-dominated PDEs like ours to present numerical instabilities at so called shocks, which are discontinuities in the solution\cite{deriveENO}. The Lax-Wendroff method we use presents this problem and rapidly growing instabilities appear at shocks independently of $\Delta t$ and $\Delta V$. As observed both from our pseudo-analytic and numerical analysis (see sections \ref{stationary} and \ref{stability}), in the problem we study here these discontinuities do not arise in the whole parameter space. We restrict the numerical analysis to the regions where they do not arise and our numerical approach gives non-oscillatory results. Essentially Non-Oscillatory (ENO) and High-Order Accurate Weighted Essentially Non-Oscillatory (WENO) schemes prevent the appearance of these spurious oscillations\cite{deriveENO, efficientENO, scholarpediaWENO}, and they could allow for a numerical study of our system in the remaining region of the parameter space. This is left for a future work.






\section{Invariant distributions}
\label{stationary}


In this section we investigate the invariant membrane potential distributions $p_s(V)$, i.e. the stationary solutions to equation \ref{diffEq} with constraints \ref{pnxDef} and \ref{norm}. It is important to note that all the solutions given here must be unstable in the finite $N$ case, where activity will cease after a finite amount of time\cite{robertTouboul, drogoulVeltz}.

For power-law spiking rates of the form $\phi(V)=(\gamma V)^n$, with $\phi(0)=0$, the trivial invariant distribution $p_s(V)=\delta(V)$ is always solution to the equation when assuming a stationary density $p(V,t)=p_s(V)$\cite{fournierLocherbach, robertTouboul, drogoulVeltz}. This state will be called \emph{death} since all neurons have zero membrane potential and no activity is present in the network ($\rho=0$). However, besides this trivial solution, a second non-trivial distribution also solves the equation. For this choice of $\phi(V)$ it reads
\begin{eqnarray}
p_s(V) = \frac{ p_{s,0} }{1-V/V_c} \exp{ \left( - \int_0^V{ \frac{ \phi(V') - (\lambda+\alpha) }{ W \rho_s + \lambda \bar{V}_s - (\lambda+\alpha)V' } {\rm d}V'} \right) } \mathds{1}_{V<V_c} = \nonumber \\
= \frac{ p_{s,0} }{ 1-V/V_c } \exp{ \left( - C \frac{(V/V_c)^{n+1}}{n+1} F_{(2,1)}(1, n+1; n+2, V/V_c) \right) } \mathds{1}_{V<V_c},
\label{genSol}
\end{eqnarray}
with
\begin{eqnarray}
p_{s,0} = \frac{\rho_s}{\lambda \bar{V}_s + W \rho_s}, \hspace{1 cm}
V_c = \frac{\lambda \bar{V}_s + W \rho_s}{\lambda+\alpha}, \label{defXcnRhos0} \hspace{1 cm}
C = \frac{(\gamma V_c)^n}{\lambda + \alpha} = \frac{\phi(V_c)}{\lambda+\alpha} \mbox{ and } \label{defC} \\
F_{(2,1)}(1, n+1; n+2; y) = (n+1) \int_0^1{ \frac{t^n}{1-t z} {\rm d}t }
\end{eqnarray}
is the hypergeometric function defined from Gauss's hypergeometric series, and $\mathds{1}_{V<V_c}$ is the indicator function, $1$ where the condition is fulfilled and $0$ otherwise. This solution, however, is restricted to a finite sub-space of the parameter space. In particular to that where exist constants $\rho_s, \bar{V}_s \in \mathbb{R}^+$ such that conditions \ref{pnxDef} and \ref{norm} are fulfilled. In \cite{fournierLocherbach} it was proven for $\alpha=0$ that the condition from the definition of $\rho$ is always fulfilled. We checked this to be true for $\alpha>0$ and $\phi(V)$ of the above given form. Then, $\bar{V}_s$ and $p_s$ need to be determined self-consistently from the constraint defined by the definition of $\bar{V}$ in equation \ref{pnxDef} and from the normalization condition in \ref{norm}. In general, this function does not have analytic integral, which can only be obtained for $n=1$, where an alternative expression for $p_s(V)$ can be found (see below). The impossibility to get an analytic expression for the integrals in conditions \ref{pnxDef} and \ref{norm} prevents us from obtaining closed expressions for the parameters $\rho_s$ and $\bar{V}_s$. For this reason we perform the integrals computationally.



Here $p_{s,0}$ is the value at $V=0$, so that the stationary solution \ref{genSol} fulfills the boundary condition in equation \ref{extConds}, $V_c$ is the largest membrane potential found on the population, i.e. $p_s(V)$ is of compact support $V \in \big[0, V_c)$ with $V_c \rightarrow \infty$ as $\lambda+\alpha \rightarrow 0$, and $C$ is the single parameter that determines the different regimes of the system. In particular,
\begin{equation}
\lim_{V \rightarrow V_c^-}{\frac{p_s(V)}{p_{s,0}}} = \begin{cases} 0 & \mbox{ for } C>1 \\ K(n) & \mbox{ for } C=1 \\ \infty & \mbox{ for } C<1, \end{cases}
\end{equation}
where $K(n)$ are finite non-zero constants that only depend on $n$. Therefore, $p_s(V)$ is continuous at the critical $V$, $V_c$, for $C>1$ while it presents there a finite and an infinite discontinuity, respectively, for $C=1$ and $C<1$. Note that the critical value $C=1$ is attained when spiking rate at the critical point $\phi(V_c)$ equals the combined pull of leakage and gap junctions, $\lambda+\alpha$. When the latter dominates, an infinite accumulation of neurons occurs in a small range of membrane potentials near $V_c$, while one finds neurons with $V>V_c$ with zero probability. Interestingly this same behavior is observed for any $n \in \mathbb{N}$, including the linear case with $n=1$.



\begin{figure}
\centering
\begin{center}
\begin{tabular}{c}
\includegraphics[width=0.85 \textwidth]{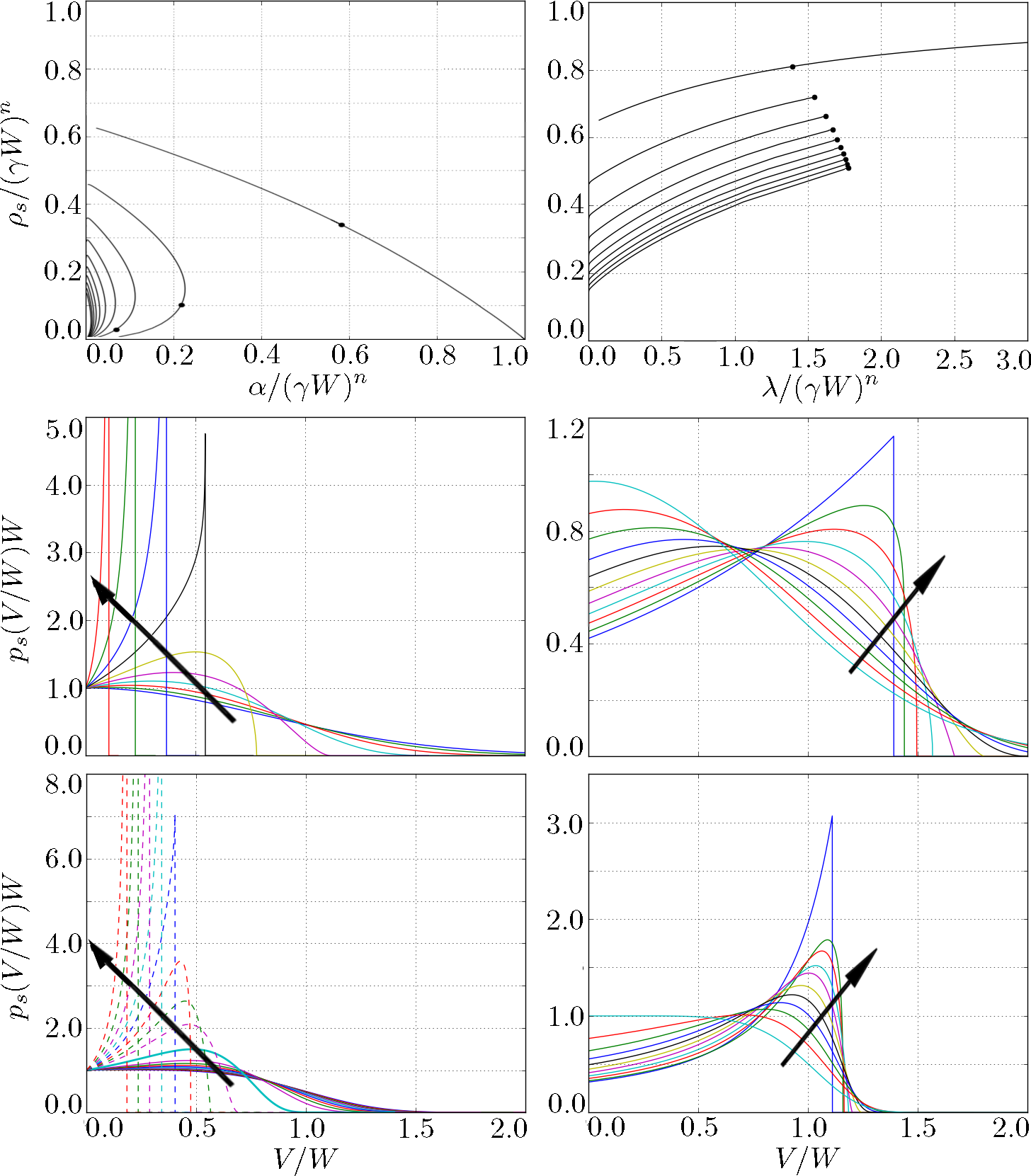}
\end{tabular}
\captionof{figure}{Normalized fraction of spiking neurons (top row) and normalized probability density functions $p_s(V)$ of the distribution of membrane potentials $V$ (mid and bottom rows) in the steady state of a network of $N \rightarrow \infty$ neurons connected all-to-all by chemical synapses of strength $W/N$ which fire stochastically with a spiking rate probability $\phi(V)= (\gamma V)^n$. The neurons also undergo a leakage of ions with the extracellular medium at a rate $\alpha$ (left column) or with the intracellular medium of all other neurons at a rate $\lambda$, through gap junctions or electrical synapses (right column). Different powers $n=1, ..., 10$ (from top to bottom) are considered for the top row. Different values of $\alpha$, changing from top to bottom along the corresponding $\rho_s{\alpha}$ curves (top left figure) in the direction of the arrows, are considered both for the $n=1$ (left mid) and $n=2$ (left bottom) cases. The distributions in solid lines sit on the upper branches of $\rho_s(\alpha)$ and are expected to be stable, while those in dashed lines are on the lower branch and are probably unstable \cite{robertTouboul}. For the cases with gap junctions in the right column, different values of $\lambda$ are considered both for the mid, with $n=1$, and the bottom, with $n=5$, figures. In this case the arrows point in the direction of increasing $\lambda$ along the $\rho_s(\lambda)$ curve (top right figure). All distributions are of compact support $\big[0, V_c)$ and the discontinuity at $V=V_c$ appears in all cases for $\phi(V_c)/(\lambda+\alpha)=1$ (at the black dots in the $\rho_s$ curves).}
\label{FigLGSep}
\end{center}
\end{figure}

\begin{figure}
\centering
\begin{center}
\begin{tabular}{c}
\includegraphics[width=0.6 \textwidth]{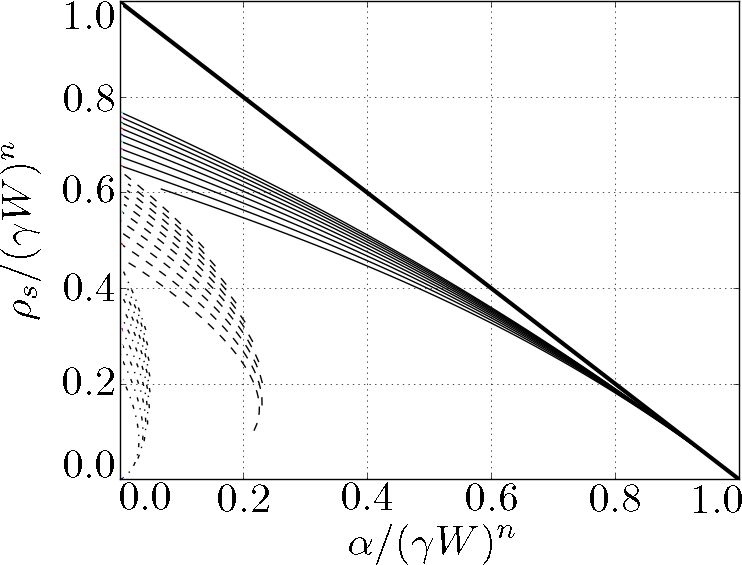}
\end{tabular}
\captionof{figure}{Normalized fraction of spiking neurons per unit time, $\rho_s/(\gamma W)^n$ for the same network as in figure \ref{FigLGSep} in the stationary regime. $\rho_s/(\gamma W)^n$ is plotted as a function of normalized leakage intensity $\alpha/(\gamma W)^n$, with spiking rate probability $\phi(V)$ with different powers $n=1$ (solid lines), $2$ (dashed) and $6$ (dot-dashed), and different gap-junction strengths $\lambda$ ($0.0$, $0.1(\gamma W)^n$, ..., $0.9(\gamma W)^n$, from bottom to top) in each case. The limit $\lim_{\lambda \rightarrow \infty}{\rho_s/(\gamma W)^n} = (1-\alpha/(\gamma W)^n)$ for $n=1$ is plotted in black solid thick line.}
\label{FigLG}
\end{center}
\end{figure}

\paragraph{Linear power firing rate (n=1):} When $n=1$ the above expression can also be written as
\begin{eqnarray}
p_s(V) = p_{s,0} (1-V/V_c)^{-1+C} e^{\frac{C V}{V_c}},
\end{eqnarray}
with $C$ as defined in equation \ref{defC}. This equation allows one to extract $\rho_s/(\gamma W)$ and $\bar{V}_s/W$ in terms of $\alpha/(\gamma W)$ and $\lambda/(\gamma W)$ using expressions
\begin{equation}
\int_0^\infty{p_s(V) {\rm d}V} = \left( \frac{\lambda+\alpha}{\lambda+W \gamma} \right) C^{1-C} e^C \left[ \Gamma(C)-\Gamma(C,C) \right] \equiv \left( \frac{\lambda+\alpha}{\lambda+W \gamma} \right) f(C) = 1
\label{normn1}
\end{equation}
and $p_s/(\gamma W) = \bar{V}_s/W$, which gives $C = \rho_s (\lambda+\gamma W)/(\lambda+\alpha)^2$. Here $\Gamma(x)$ and $\Gamma(x,y)$ are the complete and incomplete \emph{gamma} functions, respectively. From the limits $f(C) \rightarrow 1$ for $C \rightarrow 0$ and $f(C) \rightarrow \infty$ for $C \rightarrow \infty$, one finds that equation \ref{normn1} only has solution for $\alpha/(\gamma W) \in (0,1)$, which was shown to be stable for $\lambda=0$ in \cite{robertTouboul}. Here we see how this is the region where activity persists independently of $\lambda$. This means that the network activity will always die out when $\alpha \geq (\gamma W)$ and gap junctions will not be able to rescue it. Solving numerically this equation for $C$ one finds the dependence of $\rho_s/(\gamma W)$ on $\alpha/(\gamma W)$ and $\lambda/(\gamma W)$ (see figures \ref{FigLGSep} and \ref{FigLG}). Note how activity decreases monotonically for increasing $\alpha$ and increases with $\lambda$, attaining a maximal value of $\rho_s/(\gamma W) = (1-\alpha/(\gamma W))$ when $\lambda \rightarrow \infty$.




\paragraph{Neurons with no leakage connected by chemical but not electrical synapses ($\lambda=\alpha=0$):} When no leakage nor gap junctions are present the non-trivial invariant distribution takes on the simple expression \begin{eqnarray}
p_s(V) = \frac{1}{W} \exp{ \left( - \frac{ \gamma^n V^{n+1} }{ W p_s (n+1) } \right) }, \mbox{ where } \hspace{.5 cm}
\frac{p_s}{(W \gamma)^n} = \frac{1}{(n+1) \Gamma\left( \frac{n+2}{n+1} \right)^{n+1}}
\label{SolNLNG}
\end{eqnarray}
is obtained from the normalization condition \ref{norm}, which fully determines the solution since $\bar{V}_s$ does not appear in $p_s(V)$. The normalized activity $\rho_s/(\gamma W)^n$ is maximal for n=1, where the distribution is a Gaussian centered at zero, and it decays as $\approx 1/(n+1)$ in the deterministic limit $n \rightarrow \infty$, where the distribution approaches a step function of membrane potentials uniformly distributed across all sub-threshold values. In \cite{fournierLocherbach} the convergence to and stability of this solution were proven under certain conditions.


For $n>1$ and $\lambda+\alpha>0$ we cannot get an expression for the integral of $p_s(V)$, which prevents us from obtaining an analytic expression for the critical $\alpha$. Moreover, its numerical computation becomes hard for $C<1$ due to the divergence at $V=V_c$. For this reason, in most of the cases we compute $p_s$ versus $\alpha$ and/or $\lambda$ from the integrals computed numerically and only in the range $C \geq 1$.

When the neurons are leaky but they are just connected by chemical synapses, i.e. $\alpha>0$ and $\lambda=0$, $\bar{V}_s$ does not appear in the solution while $\rho_s$ can still be determined from the normalization condition. It is convenient to parametrize the functions on $C$, which gives
\begin{eqnarray}
\frac{\alpha}{(\gamma W)^n} = \frac{1}{C g(C;n)^n}, && \frac{\rho_s}{(\gamma W)^n} = \frac{1}{C g(C;n)^{n+1}},
\end{eqnarray}
with
\begin{equation}
g(C;n) = \int_0^1{ \left( \frac{p_s(y)}{p_{s,0}} \right) {\rm d}y }
\label{gdef}
\end{equation}
and $y=V/V_c$. The convergence of the integral is hard to determine for $C<1$, where we used the \emph{Computational knowledge engine} from \emph{WolframAlpha}\textsuperscript{\textregistered} for $n=2, 3$ (see figure \ref{FigLGSep}). Hence, as shown in \cite{robertTouboul}, when $\alpha/(\gamma W)^n<\alpha_{Nc}(n)$ for some critical values $\alpha_{Nc}$, two non-trivial invariant distributions solve equation \ref{diffEq} besides the trivial \emph{dead} state with $p_s(V)=\delta(V)$. We expect $\rho_s \rightarrow 0$ as $\alpha \rightarrow 0$ along the lower branch when $n>1$. The two non-\emph{dead} distributions for any given $\alpha/(\gamma W)^n<\alpha_{Nc}$ have different $\rho_s$, and in \cite{robertTouboul} computational simulations showed how only that with higher activity is stable. Similar results were obtained in \cite{brochiniDiscGL} for the discrete-time case. The lower branch, thus, seemed to belong to the separatrix of the basins of attraction of the two stable solutions. This suggests the presence of a saddle-node bifurcation at $\alpha_{Nc}$, where the stable and unstable fixed points meet and stability is lost. It also suggests distributions that are discontinuous at $V=V_c$ to be stable only for $n=1$. Computations performed using the numerical method described in \ref{methods} gave stable invariant distributions for all considered $\alpha$ and $\rho_s$ and long time periods $T$ (see Section \ref{stability}). This does not allow to draw stronger conclusions on the stability of the system. For this it is more convenient to study the linearized system close to the fixed point, which is left for a future work.

We note how $\alpha_{Nc}(n)$ decreases monotonically with $n$, which suggests its convergence to $0$ in the deterministic limit, i.e. as $n \rightarrow \infty$, and the only invariant distribution to be the \emph{dead} state.
When the neurons are connected by gap junctions besides the chemical synapses, $\lambda>0$ and $\bar{V}_s$ enters the solution. Thus, we also need the condition defining $\bar{V}$ in \ref{pnxDef} in order to fully determine the solution. In particular, the parametrized curves read
\begin{eqnarray}
\frac{\lambda}{(\gamma W)^n} = \frac{\lambda_\textrm{rel}}{C} \left( \frac{ 1 }{ g(C;n) - \lambda_\textrm{rel} h(C;n) } \right)^n, \\
\frac{\bar{V}_{s}}{W} = \frac{ h(C;n)/g(C;n) }{ g(C;n) - \lambda_\textrm{rel} h(C;n) } \hspace{0.5cm} \mbox{ and } \hspace{0.5cm} \frac{\rho_{s}}{(\gamma W)^n} = \frac{1}{C g(C;n)} \left( \frac{ 1 }{ g(C;n) - \lambda_\textrm{rel} h(C;n) } \right)^n,
\end{eqnarray}
where $g$ is as defined in equation \ref{gdef},
\begin{equation}
h(C;n) = \int_0^1{ y \left( \frac{p_s(y)}{p_{s,0}} \right) {\rm d}y }\mbox{ and we have defined } \hspace{0.5cm} \lambda_\textrm{rel} = \frac{\lambda}{\lambda+\alpha}.
\label{hdef}
\end{equation}
For both $\lambda>0$ and $\alpha>0$, the curves with fixed $\lambda$ or $\alpha$ are obtained from computing $\lambda_\textrm{rel}(C)$ from the above equations. The leakage case is recovered when $\lambda_\textrm{rel} \rightarrow 0$ and that with gap junctions and no leakage when $\lambda_\textrm{rel} \rightarrow 1$.

We show in figure \ref{FigLGSep}(top-right) the normalized activity $\rho_s/(\gamma W)^n$ as a function of normalized gap-junction strength $\lambda/(\gamma W)^n$ for $\alpha=0$ and different powers $n=1, 2, 3, ..., 10$. We note how the network activity is enhanced by the gap-junctions, reaching a plateau, where $\rho_s$ is maximum, as $\lambda \rightarrow \infty$. Though this cannot be clearly seen for $n>1$, where we could only compute the integral down to $C=1$, we expect the same behavior as for $n=1$, where this convergence is found (not shown in figure).


In \cite{drogoulVeltz} the authors gave numerical evidence for a Hopf bifurcation occurring for $n=8$ and $\lambda \approx 0.25$ (assuming $\gamma=W=1$), which we also observe in our numerical simulations, as shown in the next section. The oscillations seemed to cease $\forall n<7$ within the range of computed values of $\lambda$. Therefore, a limit cycle seems to appear for some $n>1$ and positive $\lambda$. The numerical method we developed does not allow for a systematic study of the critical $\lambda$ where the bifurcation occurs and this is left for a future work.

When $\alpha>0$ the gap junctions also enhance the activity, with a greater enhancement for larger $\lambda$, as observed for $n=1$ (see figure \ref{FigLG}). Also in these cases when $n>1$ we expect $\rho_s \rightarrow 0$ as $\alpha \rightarrow 0$, independently on $\lambda$. Moreover, we also found $\alpha_{Nc}$ to be slightly enhanced by gap junctions when $n>1$, reaching a maximum value for an intermediate $\lambda$. Now the question arises of whether the oscillations found in \cite{drogoulVeltz} are also present when $\alpha>0$, and how the leakage affects them. In the next section we study this regime.



The invariant distributions present a very similar shape for all cases with or without leakage and gap junctions, as described above. The limiting distributions for very large leakage or gap junction strength are though, very different. In particular, $p_s(V) \rightarrow \delta(V)$ as $C \rightarrow 0$ for $\lambda=0$ (which implies $\alpha \rightarrow \gamma W$ if $n=1$ and $\alpha \rightarrow 0$ if $n>1$) while $p_s(V) \rightarrow \delta(V-V_c)$ in the same limit for $\alpha=0$ (which implies $\lambda \rightarrow \infty$). The limits $\lim_{\lambda \rightarrow \infty}( V_c ) = \lim_{\lambda \rightarrow \infty}( \bar{V}_s ) > 0$ and $\lim_{\lambda \rightarrow \infty} \rho_s = \phi( \lim_{\lambda \rightarrow \infty}( V_c ) ) > 0$ could not be determined.


\section{Stability of the invariant distributions}
\label{stability}

In the previous section we gave a detailed analysis of the invariant distributions of the system, but we could not investigate their stability. As discussed in the previous sections, previous studies pursued this aim, for $\lambda=0$ and $\alpha>0$ mathematically \cite{robertTouboul}, and for $\alpha=0$ but $\lambda>0$ mathematically \cite{fournierLocherbach} and numerically \cite{drogoulVeltz}. In \cite{fournierLocherbach} the convergence to the non-trivial distribution given by \ref{SolNLNG} was proven under some assumptions for arbitrary $n \geq 2$ when $\lambda=\alpha=0$.

\begin{figure}[hb]
\centering
\begin{center}
\begin{tabular}{c}
\includegraphics[width=0.98 \textwidth]{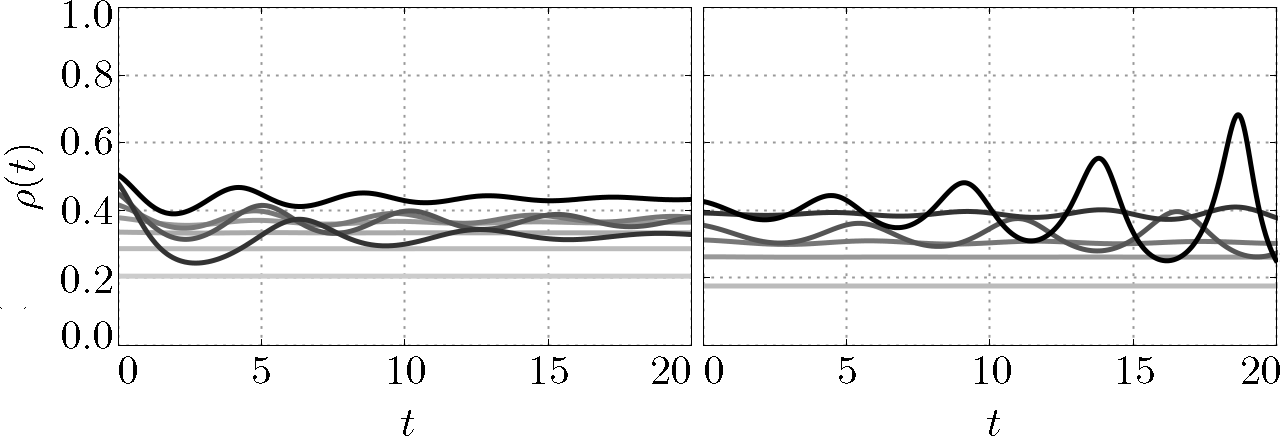}
\end{tabular}
\captionof{figure}{Numerically computed time evolution of the activity $\rho(t)$ for the same neural network of figure \ref{FigLGSep} with $\alpha=0.0$, $\gamma=W=1.0$, different gap junction strengths $\lambda$ and power $n=6$ (left) and $n=7$ (right) for the spiking rate function $\phi(V)$. $\lambda$ increases from light gray to black in the intervals $\lambda \in (6.9 \cdot 10^{-4}, 1.2)$ and $\lambda \in (5.2 \cdot 10^{-4}, 0.87)$ for the left and right figures, respectively. In all cases the non-trivial invariant distribution (see section \ref{stationary}) is used for the initial distribution, i.e. $p_0(V)=p_s(V)$.}
\label{FigGNum}
\end{center}
\end{figure}

We used the method described in section \ref{methods} to compute the time evolution of $p(V,t)$ for the different cases, with and/or without leakage, and different powers $n$, using the invariant distributions given in the previous section as initial conditions $p_0(V)$ in order to test their stability. We did this for the network with $W = \gamma = 1.0$ with and without leakage and gap junctions.



As discussed in the previous section the stability for the case with $\lambda=0$ and $n=1$ within the whole range of $\alpha/(\gamma W)^n$ values was proven in \cite{robertTouboul} while only numerical evidences were given for the $n>1$ case. The trivial invariant distribution $p_s(V)=\delta(V)$ was proven to be always stable for $n>1$. Moreover, they gave numerical evidence for the stability of the non-trivial distribution with higher activity, $\rho_s$, while that with lower $\rho_s$ appeared unstable. They concluded the latter distributions to belong to the separatrix between the basins of attraction of the other two invariant distributions. This suggests the occurrence of a saddle-node bifurcation at the critical leakage $\alpha_{Nc}$, where the stable and the unstable branches meet and stability is lost. We computed numerically the evolution of the system up to $t=50.0$ and found it to remain stable at the invariant distribution for all cases with $n$ up to $7$ and $\rho_s/(\gamma W)^n$ as low as $0.0013$. These results seem contradicting with those in \cite{robertTouboul} despite of not presenting instabilities in most of the cases, but theirs are more reliable since they considered a broader set of initial distributions around the invariant ones. Thus, we might need to apply some perturbation or wait for longer times in order to observe them. A more proper analysis would be conducted through a study of the linearized system near the fixed point, but this is left for a future work.


For the network connected by both chemical and electrical synapses with no leakage, where $\alpha=0$ and $\lambda>0$, we could recover the oscillations found in \cite{drogoulVeltz}, as we show in figure \ref{FigGNum}. We restricted the computations to $\lambda$ as large as $\lambda(C=1)$, where the discontinuity at $V_c$ appears. For this range of $\lambda$ values we could observe the oscillations to grow and probably remain stable for $n$ as low as $7$ and never for lower powers of the firing rate function. However, damped oscillations appear for $n=6$ and probably also for lower $n$. In \cite{drogoulVeltz} oscillations were never found at discontinuous distributions. This altogether with our findings, suggests them to be stable only for $n \geq 7$.

\begin{figure}[ht]
\centering
\begin{center}
\begin{tabular}{cc}
\includegraphics[width=0.98 \textwidth]{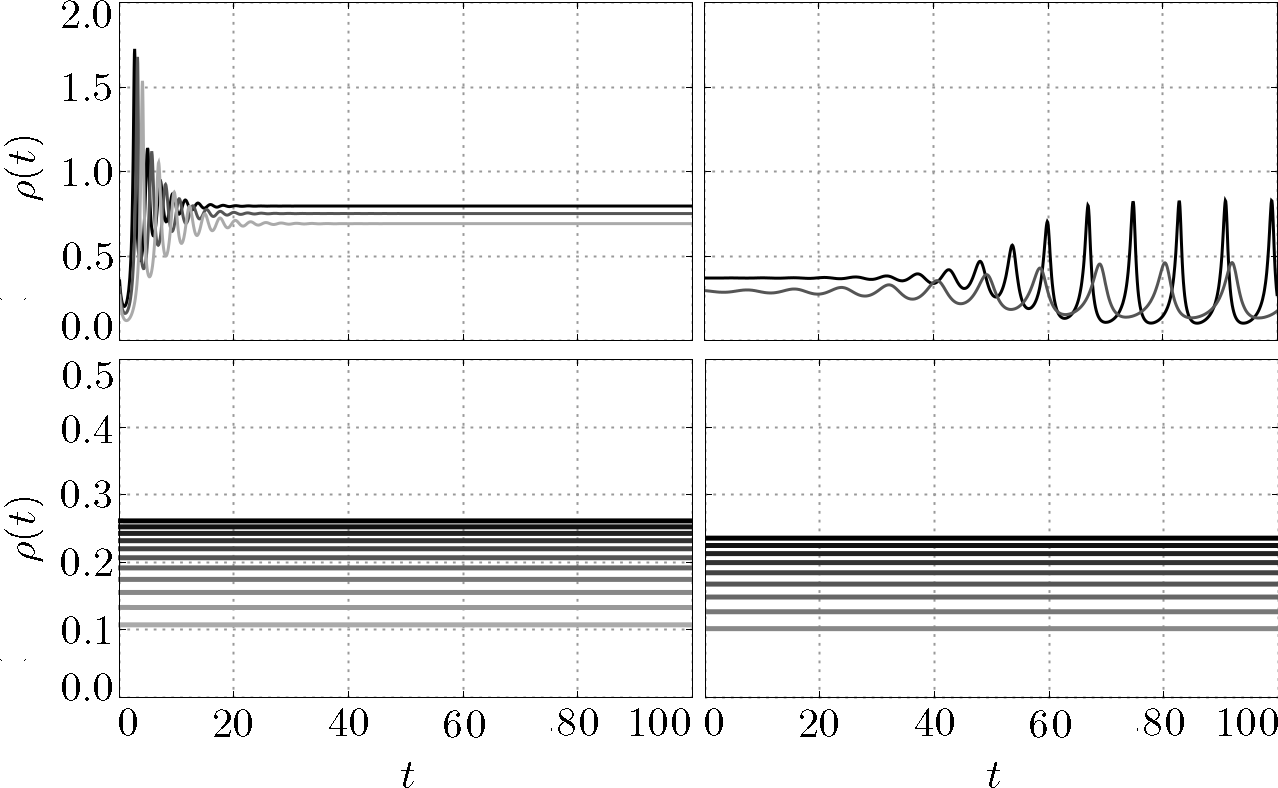}
\end{tabular}
\captionof{figure}{Time evolution of the activity $\rho(t)$ for a similar network as in figure \ref{FigGNum} but composed of neurons connected all-to-all by gap junctions of strength $\lambda=0.9$ (upper row) and $0.1$ (lower row), with a spiking rate function of power $n=6$ (left column) and $7$ (right column), which undergo a leakage of ions with the extracellular space at rates $\alpha$ increasing from light gray to black within the intervals $(4.3 \cdot 10^{-3}, 4.3 \cdot 10^{-2})$, $(1.9 \cdot 10^{-2},3.6 \cdot 10^{-2})$, $(8.0 \cdot 10^{-4}, 3.8 \cdot 10^{-2})$ and $(8.7 \cdot 10^{-4}, 3.0 \cdot 10^{-2})$ for figures from left to right and from top to bottom.}
\label{FigLGNum}
\end{center}
\end{figure}

One interesting question is whether the oscillations can still persist when leakage is added to the neurons, or whether this should make the activity to fade away and the \emph{dead} state to be the only stable one. Computations performed for leaky neurons connected both by chemical and electrical synapses, i.e. $\alpha, \lambda$ and $W$ positive, showed how this is the case,
as shown in figure \ref{FigLGNum}. In this case, a low value of $\lambda = 0.1$ gave stable distributions for $\alpha \geq \alpha_{Nc}$ and spurious oscillations arose at the lower branch of $\rho_s(\alpha)$ (see section \ref{methods}). Thus, we cannot take any conclusions about the evolution of the system in this region of the parameter space. For a large $\lambda$ of $0.9$ the range of values for $\alpha$ where $C>1$ is very narrow. Within this narrow range, again damped oscillations appeared for $n=6$, where the system reached a steady state, while oscillations grew and stabilized at a fixed amplitude and frequency for $n=7$ (see figure \ref{FigLGNum}). Thus, stable oscillations seem to appear also for $n \geq 7$ when $\alpha>0$ so that, as one could have expected, a weak leakage cannot destroy their stability. For higher leakage again one would expect the activity to \emph{die} out, but other numerical methods capable of stabilizing the spurious oscillations at shocks will be necessary to test that hypotheses. If $\lambda_c(n)$ is the critical $\lambda$ were oscillations appear for $\alpha=0$, it would be interesting to consider intermediate values of $\lambda \leq \lambda_c$ for some $n \geq 7$ to see whether $\alpha>0$ can make the stationary distribution to lose stability and stable oscillations to appear.

\section{Conclusions}
\label{conclusions}

We extended previous works on a model for networks of stochastic point spiking neurons in the mean-field limit. In particular, we gave general pseudo-analytic expressions for the non-trivial invariant distributions of membrane potentials across the network. The trivial invariant distribution $\delta(V)$ is always solution to the mean-field equations, which corresponds to the \emph{dead} state where the network presents no activity and all neurons have zero membrane potential. The non-trivial distributions, instead, are characterized by the persistence of electrical activity in the network. When stable, these states are characterized by the asynchronous activity of neurons, where they present low correlations in the spike times. We considered power law spiking rate functions and uniform leakage and all-to-all chemical and electrical synaptic strengths. These distributions are of compact support, i.e. are positive only within a finite range of membrane potentials, and present an infinite discontinuity when the combined pull of leakage and gap junctions overcomes the spiking rate at maximum potential. However, these discontinuous distributions may always be unstable.

Gap junctions always enhance the network activity. The stronger the gap junctions are the larger the enhancement, with a convergence to a $\delta(V-V_c)$ distribution, with some $V_c>0$ that we could not determine, when the strength of the gap junctions tends to infinity.

Extending previous studies, our numerical simulations show how the network can sustain oscillations of its global activity if gap junctions are permeable enough also in the presence of leakage. In these oscillatory states the neurons in the network present a high level of synchrony in their activity, in contrast to asynchronous states. The model, thus, presents a rich phenomenology despite of its mathematical simplicity, offering a promising tool for large-scale computational simulations of neural networks.


These results also suggest an important role for gap junctions in the appearance of highly synchronized states in populations of excitatory neurons, which could play a role in the activity-dependent phase of developing nervous systems. The oscillations arise precisely for power-law firing rates with high power, which give a better fit of neural activity observed experimentally.




\section*{Acknowledgements}

This work was produced as part of the USP project \emph{Mathematics, computation, language and the brain} and the FAPESP (S\~ao Paulo Research Foundation) project \emph{Research, Innovation and Dissemination Center for Neuromathematics} (grant 2013/07699-0) and is supported by the FAPESP grant 2015/10785-0.

I thank Professor Pierre Collet, from the \emph{Centre de Physique Th{\'e}orique} from the \emph{{\'E}cole Polytechnique} in France, for very fruitful discussions.




\end{document}